\documentclass[
prl,aps,
reprint,
a4paper,
superscriptaddress,
longbibliography,
floatfix
]{revtex4-1}
\usepackage{graphicx}
\usepackage{epsfig}
\usepackage{amsfonts}
\usepackage{amsmath}
\usepackage{amssymb}
\usepackage{color}
\usepackage[colorlinks=true,linkcolor=blue,citecolor=magenta,urlcolor=blue]{hyperref}

\begin{document}


\title{Bell nonlocality between sequential pairs of observers}


\author{Ad\'an Cabello}
\email{adan@us.es}
\affiliation{Departamento de F\'{\i}sica Aplicada II, Universidad de Sevilla, E-41012 Sevilla, Spain}
\affiliation{Instituto Carlos~I de F\'{\i}sica Te\'orica y Computacional, Universidad de
Sevilla, E-41012 Sevilla, Spain}


\begin{abstract}
We show that it is possible to have arbitrarily long sequences of Alices and Bobs so {\em every} (Alice, Bob) pair violates a Bell inequality. We propose an experiment to observe this effect with two Alices and two Bobs. 
\end{abstract}


\maketitle


{\em Introduction.---}Bell nonlocality, that is, the impossibility of simulating quantum mechanics with local hidden-variable models \cite{Bell64,CHSH69,BCPSW14}, is one of the most fascinating predictions of quantum mechanics and underpins many applications, including, entanglement-based quantum cryptography \cite{GRTZ02}, certified randomness \cite{AM16}, self-testing of quantum systems \cite{SB20}, and quantum advantage in shallow circuits \cite{BGK18}.

Surprisingly, after decades of research in Bell nonlocality, we still do not have the answer to some fundamental questions. Here, we focus on one of them, originally proposed by Silva {\em et al.}~\cite{SGGP15}: Is it possible to have an arbitrarily long sequence of pairs of Alices and Bobs that violate a Bell inequality?
 
\begin{figure*}[t!]\centering
	\includegraphics[width=1\textwidth]{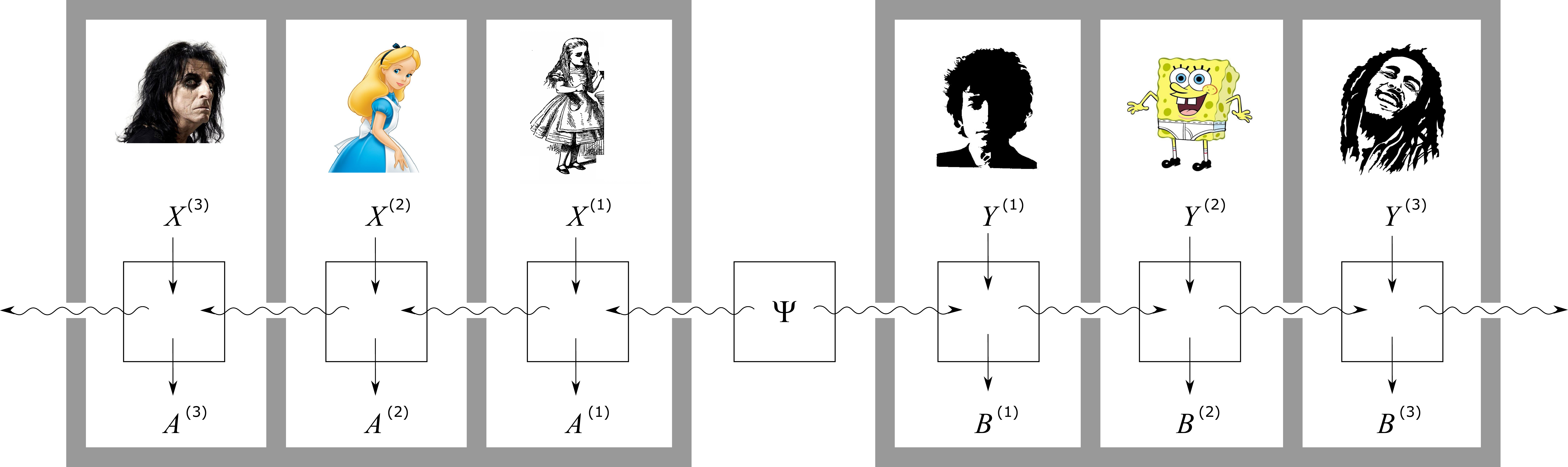}
	\caption{Three independent Alices and three independent Bobs sharing a pair of particles initially prepared in state $\Psi$ and performing, sequentially, measurements, denoted $X^{(j)}$ and $Y^{(k)}$, respectively, on each of the particles. Measurement results are denoted $A^{(j)}$ and $B^{(k)}$, respectively.}\label{Fig1}
\end{figure*}

Consider that each of Alice$^{(1)}$ and Bob$^{(1)}$ has one of a pair of entangled particles. Alice$^{(1)}$ measures her particle and then passes it to Alice$^{(2)}$ who, acting independently and without knowing the results of Alice$^{(1)}$, measures the particle and then passes it to Alice$^{(3)}$ and so on. Similarly, in a far away region, Bob$^{(1)}$ measures his particle and then passes it to Bob$^{(2)}$ who, acting independently and without knowing the results of Bob$^{(1)}$, measures it and then passes the particle to Bob$^{(3)}$ and so on, as illustrated in Fig.~\ref{Fig1}. Suppose that, for every pair $(j,k)$, the measurement choices of Alice$^{(j)}$ are spacelike separated from the results of Bob$^{(k)}$, and the results of Alice$^{(j)}$ are spacelike separated from the measurement choices of Bob$^{(k)}$. Is it possible to have arbitrarily long sequences of Alices and Bobs so {\em every} (Alice$^{(j)}$, Bob$^{(k)}$) pair violates a Bell inequality? 


Following in Silva {\em et al.}'s footsteps \cite{SGGP15,CJAHWA17,TC18}, Brown and Colbeck \cite{BC20} proved that the number of Bobs can be arbitrarily large if the number of Alices is $1$, but left open the question of whether the number of Alices can be also arbitrarily large.

Recently, Cheng, Liu, and Hall \cite{CLH21} have shown that, if the pair of particles only carry qubit-qubit entanglement, then the number of Alices cannot be larger than~$1$ when the number of Bobs is arbitrarily large if the parties are testing the Clauser-Horne-Shimony-Holt (CHSH) \cite{CHSH69} Bell inequality, but left open the question of whether the number of Alices can be also arbitrarily large for other Bell inequalities. See also a related work by Zhu {\em et al.}~\cite{Zhu21}. 
 
Here, by building on the protocol of Brown and Colbeck \cite{BC20}, we show that both the number of Alices and Bobs can be arbitrary large if the initial pair of particles carries ququart-ququart entanglement and we consider a Bell inequality with more settings and outcomes. A ququart is a four-dimensional quantum system.
Hence, we answer positively to Silva {\em et al.}'s question. 

From the fundamental side, this result implies that the same number of entangled pairs needed to produce Bell nonlocality between Alice$^{(1)}$ and Bob$^{(1)}$, a number that can be as low as $1$ \cite{AHQ20}, can be used to produce Bell nonlocality between any possible pair of spacelike separated Alice$^{(j)}$ and Bob$^{(k)}$ over spacetime. That is, a finite amount of a quantum resource (a number of entangled pairs) can produce nonclassical effects over an unbounded number of pairs of causally unconnected spacetime regions. 

However, from the applications' perspective, it is important to remark that the average violation of the Bell inequalities is small and decreases as the number of Alices and Bobs grows. Therefore, the result is not very useful in practice for most existing applications. However, new applications could exploit this effect, at least in the case of two Alices and two Bobs.


{\em Method.---}Let us consider two ququarts $A$ and $B$ in the maximally entangled state
\begin{equation}
	\label{state}
	|\Psi\rangle_{A,B} = \sum_{i=1}^4 |i,i\rangle_{A,B}.
\end{equation}
Suppose that ququart $A$ goes to Alice$^{(1)}$ while ququart $B$ goes to Bob$^{(1)}$. After Alice$^{(j)}$'s measurement, $A$ is passing to Alice$^{(j+1)}$, with $j=1,\ldots, J-1$, and after Bob$^{(k)}$'s measurement, $B$ is passing to Bob$^{(k+1)}$, with $k=1,\ldots, K-1$.

For the proofs to follow, it is convenient to visualize each ququart's Hilbert space $\mathbb{C}^4$ as a two-qubit system $\mathbb{C}^2 \otimes \mathbb{C}^2$. We denote by $A1$ and $A2$ the two qubits in ququart $A$, and by $B1$ and $B2$ the two qubits in ququart $B$. With this notation, $|\Psi\rangle_{A,B} = |\phi^+\rangle_{A1,B1} \otimes |\phi^+\rangle_{A2,B2}$, where $|\phi^+\rangle_{Ai,Bi}=\frac{1}{\sqrt{2}}\left(|00\rangle + |11\rangle \right)$ is a Bell state between qubits $Ai$ and $Bi$. 

Now consider the following measurement strategy. Each Alice$^{(j)}$ chooses between four different four-outcome measurements $X^{(j)}_{m,n}$, with $m,n \in \{0,1\}$. Each of these measurements is represented by four POVM elements: $X^{(j)}_{p,q|m,n}$, with $p,q \in \{0,1\}$. POVM stands for positive operator-valued measurement. Specifically, consider the following $16$~POVM elements ($4$ of them for each of the $4$~possible measurements) for Alice$^{(j)}$:
\begin{equation}
	\label{m1}
	X^{(j)}_{p,q|m,n} = X^{(j,A1,q,n)}_{p|m} \otimes X^{(j,A2,p,m)}_{q|n},
\end{equation}
with
\begin{widetext}
\begin{subequations}
	\label{m2}
	\begin{align}
		&X^{(j,A1,q,n)}_{p|m}=\tfrac{1}{2} \left[\openone^{(A1)} + (1-m) (1-2p)\sigma_z^{(A1)}+ m (1-2p) \gamma_j^{(1)} \sigma_x^{(A1)} \right],\\
		&X^{(j,A2,p,m)}_{q|n}=\tfrac{1}{2} \left[\openone^{(A2)} + (1-2q) \cos(\theta^{(2)}) \sigma_z^{(A2)} + (1 - 2|n-q|) \sin(\theta^{(2)}) \sigma_x^{(A2)} \right],
	\end{align}
\end{subequations}
\end{widetext}
where $\sigma_z^{(A1)}$ denotes the Pauli $z$ matrix for qubit $A1$. 

Similarly, Bob$^{(k)}$ chooses between four different four-outcome measurements $Y^{(k)}_{m,n}$, each of them represented by four POVM elements. The $16$~POVM elements for Bob$^{(k)}$ are
\begin{equation}
	\label{m3}
	Y^{(k)}_{p,q|m,n} = Y^{(k,B1,q,n)}_{p|m} \otimes Y^{(k,B2,p,m)}_{q|n},
\end{equation}
with
\begin{widetext}
\begin{subequations}
	\label{m4}
	\begin{align}
		& Y^{(k,B1,q,n)}_{p|m} = \tfrac{1}{2} \left[\openone^{(B1)} + (1-2p) \cos(\theta^{(1)}) \sigma_z^{(B1)} + (1-2|m-p|) \sin(\theta^{(1)}) \sigma_x^{(B1)} \right],\\
		& Y^{(k,B2,p,m)}_{q|n} = \tfrac{1}{2} \left[\openone^{(B2)} + (1-n) (1-2q) \sigma_z^{(B2)} + n (1-2q) \gamma_k^{(2)} \sigma_x^{(B2)} \right].
	\end{align}
\end{subequations}
\end{widetext}
Therefore, for any given $J$ and $K$ (total number of Alices and Bobs, respectively), the measurements are determined by $J+K+2$ parameters: $\theta^{(1)}$, $\theta^{(2)}$, $\gamma_j^{(1)}$, and $\gamma_k^{(2)}$, with $j=1,\ldots,J$ and $k=1,\ldots,K$.

Let us adopt the following convention. Whenever Alice$^{(j)}$ measures $X^{(j)}_{mn}$ and obtains the result $(p,q)$, this is equivalent to measuring observable $X^{(j,A1,q,n)}_{m}$ obtaining result $p$, and measuring observable $X^{(j,A2,p,m)}_{n}$ obtaining the result $q$. And similarly for Bob's measurements and outcomes.

Taking this convention into account, let us consider the following Bell nonlocality witness between Alice$^{(j)}$ and Bob$^{(k)}$:
\begin{widetext}
\begin{equation}
	S^{(j,k)}= \sum_{i \in \{1,2\}} \sum_{r,s \in \{0,1\}} \sum_{t,u,v,w \in \{0,1\}} P\left(X^{(j,Ai,t,v)}_r \oplus Y^{(k,Bi,u,w)}_s =rs\right),
\end{equation}
\end{widetext}
where $P\left(X^{(j,Ai,t,v)}_r \oplus Y^{(k,Bi,u,w)}_s =rs\right)$ denotes the probability that the sum modulo~$2$ of the results of $X^{(j,Ai,t,v)}_r$ and $Y^{(k,Bi,u,w)}_s$ is equal to the product $rs$. 

For local hidden-variable theories,
\begin{equation}
	S^{(j,k)} \le 32 \times 3 = 96.
\end{equation}
This can be proven by noticing that, for every $(i,t,u,v,w)$, local hidden-variable theories satisfy $\sum_{r,s \in \{0,1\}} P\left(X^{(j,Ai,t,v)}_r \oplus Y^{(k,Bi,u,w)}_s =rs\right) \le 3$, which is the CHSH inequality \cite{CHSH69}. That is, $S^{(j,k)}$ can be seen as a sum of $32$ independent Bell nonlocality witnesses of the CHSH type.

If measurements are implemented by means of projective (sharp) measurements \cite{Neumark40} that transform the quantum state according to L\"uders' rule \cite{Luders51}, then, for state (\ref{state}) and measurements (\ref{m1})--(\ref{m4}), the expected value for $S^{(j,k)}$ is
\begin{widetext}
	\begin{align}
		S^{(j,k)} = & 2^{5-j} \left[\gamma_j^{(1)} \sin (\theta^{(1)})+\cos (\theta^{(1)}) \prod_{g=1}^{j-1} \left(1+\sqrt{1-(\gamma^{(1)}_g)^2}\right) \right] \nonumber \\
	 & + 2^{5-k} \left[\gamma_k^{(2)} \sin (\theta^{(2)})+\cos (\theta^{(2)}) \prod_{h=1}^{k-1} \left(1+\sqrt{1-(\gamma^{(2)}_h)^2}\right) \right]+64.
	\end{align}
\end{widetext}


{\em Nonlocality for unbounded Alices and Bobs.---}Now we can prove the following result. For each pair $(J,K)$, with $J,K \in \mathbb{N}$ there are values for $\theta^{(1)}$, $\theta^{(2)}$, $\gamma_j^{(1)}$, and $\gamma_k^{(2)}$, with $j=1,\ldots,J$ and $k=1,\ldots,K$, such that $S^{(j,k)} > 96$ for all pairs $(j,k)$.

The proof follows from applying to every of the 32~CHSH witnesses in $S^{(j,k)}$, two results from \cite{BC20}. Lemma~1 in \cite{BC20} points out that, for $i \in \{1,2\}$ and $\epsilon^{(i)} >0$, the sequence defined as follows: 
\begin{equation}
	\label{sec0}
\gamma_1^{(i)} (\epsilon^{(i)}, \theta^{(i)})=(1+\epsilon^{(i)}) \frac{1-\cos(\theta^{(i)})}{\sin(\theta^{(i)})},
\end{equation}
\begin{widetext}
\begin{equation}
	\label{secn}
\gamma_j^{(i)} (\epsilon^{(i)}, \theta^{(i)})= \begin{cases} 
	(1+\epsilon^{(i)}) \frac{2^{j-1}-\cos(\theta^{(i)}) \prod_{g=1}^{j-1}\left(1+\sqrt{1-(\gamma^{(i)}_g)^2}\right)}{\sin(\theta^{(i)})} 
	 & \mbox{if } \delta_{j-1} (\epsilon^{(i)}, \theta^{(i)}) \in (0,1) \\ 
	\infty & \mbox{otherwise}
\end{cases}
\end{equation}
\end{widetext}
is positive and increasing. $\gamma_j^{(i)}(\epsilon^{(i)}, \theta^{(i)})=\infty$ indicates that violation of the corresponding CHSH inequality is not possible for this $j$.

Lemma~2 in \cite{BC20} points out that, if $\epsilon^{(i)} >0$ and $\{\gamma_j^{(i)}(\epsilon^{(i)}, \theta^{(i)})\}_{j=1}^n$ is the sequence defined by Eqs.~(\ref{sec0}) and (\ref{secn}), then, for
any $n \in \mathbb{N}$, there exists some $\theta_n^{(i)} \in (0, \frac{\pi}{4}]$ such that for
all $j \leq n$ and $\theta^{(i)} \in (0, \theta_n^{(i)})$,
$\gamma_j^{(i)}(\epsilon^{(i)}, \theta^{(i)}) < 1$.

By Lemma~2, there exists some
$\theta^{(2)}(J), \theta^{(1)}(K) \in (0, \frac{\pi}{4}]$ such that $\theta^{(2)}(J) < 1$ and $\theta^{(1)}(K) < 1$. By Lemma~1, $0 < \gamma^{(2)}_1(J) < \gamma^{(2)}_2(J) < \cdots < \gamma^{(2)}_J(J) < 1$ and $0 < \gamma^{(1)}_1(K) < \gamma^{(1)}_2(K) < \cdots < \gamma^{(1)}_K(K) < 1$. These values of $\theta^{(2)}(J)$ and $\theta^{(1)}(K)$ define sequences $\{\gamma^{(2)}_i(J)\}_{i=1}^J$ and $\{\gamma^{(1)}_i(K)\}_{i=1}^K$, respectively, which, by construction, lead to violations of all the 32~CHSH inequalities for every pair $(j,k)$, which translates into a violation of $S^{(j,k)} \le 96$ for all $(j,k)$.


{\em Proposed experiment.---}Here we propose an experiment to test this result. It involves just two Alices and two Bobs. Larger numbers of Alices and Bobs would require visibilities that are difficult to achieve in practice. 

We prepare state (\ref{state}) and implement measurements (\ref{m1})--(\ref{m4}), adopting the parametrization given by Eqs.~(\ref{sec0}) and (\ref{secn}), with $\epsilon^{(1)}=\epsilon^{(2)}=0.5577$ and $\theta^{(1)} = \theta^{(2)} = \frac{\pi}{4}-0.2999$, which is a valid choice for Lemma~2 and leads to the largest violations in the case of two Alices and two Bobs. 
The resulting values for $S^{(j,k)}$ are in the table below. Each of them violates the respective Bell inequality $S^{(j,k)} \le 96$.
\begin{table}[h!]
	\begin{center}
	\begin{tabular}{l| c c}
		 & Alice$^{(1)}$ & Alice$^{(2)}$ \\
		 \hline
		Bob$^{(1)}$ & $98.06$ & $98.37$ \\
		Bob$^{(2)}$ & $98.37$ & $98.67$ 
	\end{tabular}
\end{center}
\end{table}

To illustrate why it is difficult to observe the effect with more parties, let us consider a valid choice leading to the largest violations in the case of three Alices and three Bobs: $\epsilon^{(1)}=\epsilon^{(2)}=0.27665$ and $\theta^{(1)} = \theta^{(2)} = \frac{\pi}{4}-0.6219$. The resulting values for $S^{(j,k)}$ are in the table below. Each of them violates the respective Bell inequality $S^{(j,k)} \le 96$, but the violations are too small to be observed in the laboratory.
\begin{table}[h!]
	\begin{center}
	\begin{tabular}{l| c c c}
		& Alice$^{(1)}$ & Alice$^{(2)}$ & Alice$^{(3)}$ \\
		\hline
		Bob$^{(1)}$ & $96.12$ & $96.13$ & $96.20$ \\
		Bob$^{(2)}$ & $96.13$ & $96.14$ & $96.21$ \\
		Bob$^{(3)}$ & $96.20$ & $96.21$ & $96.28$
	\end{tabular}
\end{center}
\end{table}

What is the most adequate physical system to test these predictions? Previous experiments with photons have achieved up to three sequential projective (sharp) measurements on nonentangled \cite{ARBC09} and entangled \cite{LHC16} ququarts, and up to three sequential unsharp measurements on entangled qubits \cite{FCT20}. In all these experiments, the results of the sequential measurements are all read out at the end of the sequence, when each photon is detected. This is a drawback for our purpose, as the main interest of the result is the mutual independence of the Alices and Bobs in making decisions and reading outcomes. However, photons offer visibilities $\ge 0.99$, which are difficult to obtain in other physical systems. 

Experiments with trapped ions have achieved three sequential sharp measurements on nonentangled ququarts \cite{KZG09} and sequences of millions of sharp measurements on nonentangled qutrits, each of them randomly chosen after the previous measurement is read out \cite{LMZNCAH18}. In principle, with trapped ions, it would possible to perform sequential unsharp measurements on each of a pair of entangled ququarts. The limiting factor is the visibility.

An alternative would be using, instead of ququarts, pairs of superconducting qubits \cite{Chen21} plus extra ancillary qubits for sequential nondestructive unsharp measurements on the pairs of qubits that replace the ququarts. This would allow reading out the intermediate results before the next measurements are performed. However, the visibility might be a problem.

Finally, further explorations are needed to check whether Bell nonlocality between unbounded pairs of Alices and Bobs can be achieved with simpler Bell inequalities and local quantum systems of smaller dimension, and whether other strategies may lead to violations that are easier to test experimentally.


We thank Shuming Cheng, Ehtibar Dzhafarov, Michael J. W. Hall, Lijun Liu, and Armin Tavakoli for conversations. This work has been supported by Project Qdisc
(Project No.\ US-15097), with FEDER funds, and QuantERA grant SECRET, by MINECO (Project
No.\ PCI2019-111885-2).



\end{document}